\documentclass[12pt]{iopart}

\usepackage{iopams}
\usepackage{graphicx}
\usepackage{color}
\usepackage{subfigure}

\begin{document}

\title{Emergence of Blind Areas in Information Spreading}

\author{Zi-Ke Zhang$^{1,2,\S}$, Chu-Xu Zhang$^{1,3,\S}$, Xiao-Pu Han$^{1,2}$ and Chuang Liu$^{1,2}$}

\address{$^1$ Institute of Information Economy, Hangzhou Normal University, 311121 Hangzhou, People's Republic of China\\
$^2$ Alibaba Research Center for Complexity Sciences, Hangzhou Normal University, 311121 Hangzhou, People's Republic of China\\
$^3$ Web Sciences Center, University of Electronic Science and Technology of China, 611731 Chengdu, People's Republic of China\\
$^\S$ These authors contributed equally to this work\\}

\ead{liuchuang@hznu.edu.cn}


\begin{abstract}
Recently, contagion-based (disease, information, etc.) spreading on social networks has been extensively studied.
In this paper, other than traditional full interaction, we propose a partial interaction based spreading model, considering that the informed individuals
would transmit information to only a certain fraction of their neighbors due to the transmission ability in real-world social networks.
Simulation results on three representative networks (BA, ER, WS) indicate that the spreading efficiency is highly correlated with the network heterogeneity.
In addition, a special phenomenon, namely \emph{Information Blind Areas} where the network is separated by several information-unreachable clusters,
will emerge from the spreading process. Furthermore, we also find that the size distribution of such information blind areas obeys power-law-like distribution, which has very similar exponent with that of site percolation. Detailed analyses show that the critical value is decreasing along with the network heterogeneity for the spreading process, which is complete the contrary to that of random selection. Moreover, the critical value in the latter process is also larger than that of the former for the same network. Those findings might shed some lights in in-depth understanding the effect of network properties on information spreading.
\end{abstract}

\maketitle

\section{\label{S1:Intro}Introduction}
With the advent of various online Social Networking Services (SNS), the dissemination of information through Internet, also known as \emph{word-of-mouth} or \emph{peer-to-peer} spreading, has attracted much attention for researchers in recent years \cite{Van-Mieghem-Van-de-Bovenkamp-2013-PRL,Holme-Saramaki-2012-PR}. Following the rapid development of database technology and computational power, various spreading phenomena on large-scale social networks, such as the news \cite{Chen-Chen-Gunnell-Yip-2013-PlosOne}, rumors \cite{Doer-Fouz-Friedrich-2012-CACM,Moreno-Nekovee-Pacheco-2004-PRE}, innovation \cite{Montanari-Saberi-2010-PNAS}, behavior \cite{Centola-2010-Science,Centola-2011-Science}, culture \cite{Dybiec-Mitarai-Sneppen-2012-PRE}, viral marketing \cite{Aral-2011-MS}, etc, can now be deeply studied in terms of theoretical models as well as empirical analyses.

Generally, the information spreading dynamics is usually studied under the framework of epidemic spreading \cite{Daley-Kendall-1964-Nature}. Therefore, in many studies, the process of information diffusion is regraded equally as the disease propagation, which informs agents to transmit information to their neighbors via social connections \cite{Castellano-Fortunato-Loreto-2009-RMP}. Among them, the Susceptible-Infected-Recovered (SIR) model is the most commonly used method to describe the information spreading process, where individuals would lose interest of further contribution due to a variety of less predictable factors, which is very similar to the $R$ state of the SIR model in epidemic spreading. The interplay between the spreading dynamics and network structure is a key insight in the study on network spreading dynamics \cite{Zhou-Fu-Wang-2006-PNS,Nagata-Shirayama-2012-PA}. The network structure affects both the spreading speed and prevalence through features such as the shortest path length, degree distribution, degree correlations, and so on. In addition, the upper bound of informed proportion is proved to be approximate to 80\% for the SIR model on random networks when the population size is infinite \cite{Sudbury-1985-JAP}. Previous works have also revealed that there indeed exists a propagation threshold of information spreading on the small-world network \cite{Zanette-2001-PRE,Stone-Mckay-2011-EPL}, and the spreading is much faster and broader than that on regular network for the existence of the long-range edges \cite{Keeling-Eames-2005-JRSI}. Moreno \emph{et al.} \cite{Moreno-Nekovee-Pacheco-2004-PRE,Moreno-Nekovee-Vespignani-2004-PRE} studied the rumor spreading on scale-free networks, and found that the existence of hub nodes can enhance the spreading speed rather than the influenced scope. Recently, there is a vast class of studies focusing on the spreading dynamics on interconnected networks
\cite{Gao-Buldyrev-Stanley-Havlin-2012-NP}, and there is a mixed phase below the critical infection strength in weakly coupled networks \cite{Dickison-Havlin-Stanley-2012-PRE}. Furthermore, some interesting phenomena are discovered based on the information diffusion on real social systems, such as the telephone interactions \cite{Karsai-Kivela-Pan-Kaski-Kertesz-Barabasi-Saramaki-2011-PRE,Miritello-Moro-Lara-2011-PRE}, tweets \cite{Goel-Watts-Goldstein-2012-EC,Myers-Zhu-Leskovec-2012-KDD} and emails \cite{Karsai-Kivela-Pan-Kaski-Kertesz-Barabasi-Saramaki-2011-PRE}. Empirical results indicate that the human active patterns \cite{Iribarren-Moro-2009-PRL,Doerr-Blenn-VanMieghem-2013-PLoSONE} and the role of weak ties \cite{Zhao-Wu-Xu-2010-PRE} would strongly affect the information spreading.

However, there are also plenty of researches arguing that the underlying mechanism of information diffusion should be fundamentally different from that of epidemic spreading. L$\ddot{u}$ \emph{et al.} \cite{Lu-Chen-Zhou-2011-NJP} summarized the significant differences between them, and concluded that the information, by considering the social enhancement, would spread more effectively in regular networks than that in random networks. This would to some extent support the real human experiment reported by Centola \cite{Centola-2010-Science}. Most theoretical models assumed that the informed agents would transmit the information to all their neighbors \cite{Pastor-Satorras-Vespignani-2001-PRL, SunY-2013-Arxiv} or just one randomly chosen neighbor \cite{Yang-Zhou-2012-PRE} in one single time step. However, since passing messages along would take a perceived transmission cost \cite{Banerjee1993} in real societies, the diffusion targets would be selected among individuals with potential interests \cite{Wu-Huberman-Adamic-Tyler-2004-PA}. Thus, the information flow would travel through social connections thereby depending on the properties of observed networks \cite{Iribarren-Moro-2011-PRE}.

In this paper, we propose a theoretical model with considering the effect of partial interaction, where the number of interacted neighbors is in proportion to the spreader's degree. Simulation results show that the spreading process percolates on all three classical networks (ER, BA and WS networks), which is quite similar with the site percolation in statistical physics \cite{Parshani-Carmi-Havlin-2010-PRL}. In addition, it is observed that the percolation peaks later on WS network than other two networks, which could be regarded as a good explanation of why information can spread more widely on WS network. Furthermore, it is also astonishing to find that the size distribution of unreachable clusters, namely the \emph{Information Blind Area}, exhibits the power-law property with exponent $\gamma=2.9$, which is the universal phenomenon in the community distribution \cite{Clauset-Newman-Moore-2004-PRE,Radicchi-Castellano-Cecconi-Loreto-Parisi-2004-PNAS}. 

\section{The Model}
\label{Sec:Model}
In this paper, we consider a synthetic network $G(N,E)$, where $N$ is the number of nodes and $E$ is the number of links, representing the individuals and their interactions, respectively. Analogous with the SIR model, every individual would be and only be at one of the three following states during the process of information spreading,
\begin{itemize}
\item \emph{Uninformed} (S). The individual has not yet received the information, and is analogous to the susceptible state of the SIR model;
 \item \emph{Informed} (I). The individual is aware of the information but has not transmitted it, and is analogous to the infected state in the SIR model;
 \item \emph{Exhausted} (R). After transmitting the information, the individual will probably lose interest and no longer transmit it, thus is analogous to the recovered state of the SIR model. 
\end{itemize}

Subsequently, we are mainly interested in the effect of partial interaction. That is to say, each infected node, so-called \emph{spreader} \cite{Kitsak-Gallos-Havlin-Liljeros-Muchnik-Stanley-Makse-2010-NP}, will only influence a certain fraction of its neighbors. Thus, the model can be described as follows.

\begin{itemize}
\item Initially, one arbitrary node is randomly picked as the \emph{Information Seed} (I-state) and the rest remain uninformed (S-state).
\item Then, the seed will transmit information to $\alpha$ fraction of its neighbors and then becomes exhausted (R-state), where $\alpha \in [0,1]$;
\item For simplicity, we assume that all individuals fully trust their social connections. Consequently, each individual will approve its spreader's information once she receives it;
\item After approval, she becomes an informed individual or a spreader (I-state), and will transmit the information to all $\alpha$ fraction of her neighbors and becomes exhausted (R-state);
\item The above process will repeat until there is no individual transmitting the information any more.

\end{itemize}

\begin{figure}[htb]
  \centering
  \includegraphics[width=7.5cm,height=6cm]{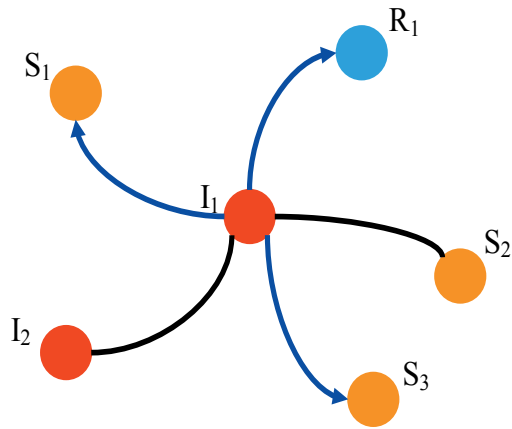}
  \caption{(Color online)\label{Fig:fig1} Illustration of the spreading rule in the proposed model with three uninformed nodes, $S_{1}$, $S_{2}$ and $S_{3}$, two informed nodes $I_{1}$ and $I_{2}$, and one exhausted node $R_{1}$. Each of them has the same probability $\frac{1}{5}$ to receive information from the informed node $I_{1}$ at each round. When $\alpha$=0.6, only $\alpha k$=3 nodes, e.g. $S_{1}$, $S_{3}$ and $R_{1}$, will receive information.}
\end{figure}

In this model, the popular individuals (nodes with large degree) tend to interact with more neighbors when they approve the information. However, since there is transmission cost in the spreading process \cite{Banerjee1993}, that is to say, individuals would not be able to interact with all their own neighbours. Therefore, we propose a tunable parameter $\alpha \in [0,1]$, representing the interaction strength. Thus, the number of neighbors, $\alpha k$, will be informed is proportional to the spreader's degree $k$, neglecting the neighbors' states. And the probability of each neighbor of being selected is $\frac{1}{k}$ at each round, and the repeating selection is prohibited. Figure \ref{Fig:fig1} shows the proposed spreading rule. By setting $\alpha$=0.6, the informed node $I_{1}$ can only transmit information to three neighbors, two uninformed nodes $S_{1}$, $S_{3}$ and one exhausted node $R_{1}$. Therefore, The present rule implies a different feature of information spreading, \emph{Partial Interaction}, which is usually neglected in the standard SIR model and its variants for information spreading. In addition, the immediate influence in the present model corresponds to the two parameters in traditional SIR model, the infected probability $\beta$ and the recovered probability $\mu$, are both equal to one in the proposed model.

To better investigate the effect of the present model, we perform analyses on three kinds of network: (i) ER network \cite{Erdos-Renyi-1959-PM}: a random graph where $N$ nodes connected by $E$ edges which are chosen randomly from all the $N(N-1)/2$ possible edges; (ii) BA network \cite{Barabasi-Albert-1999-Science}: a growing network where each newly added node connects to $m$ old nodes by preferential attachment mechanism; (iii) WS network \cite{Watts-Strogatz-1998-Nature}: Randomly reshuffle links of a regular network with probability $\gamma$ and result in the small-world network. As a consequence, we generate corresponding BA, ER and WS networks with the same network size $N$=10000, $m$=3, $\gamma$=0.5 and the average degree $\langle k \rangle $=6. In addition, to alleviate the effect of randomly selecting the spreading $seed$, all simulation results are obtained by averaging over 1000 independent realizations.

\section{Results and Analysis}
\label{Sec:Results and Discussions}
\subsection{Exhausted Rate}
\label{Sec:Exhausted Rate}

\begin{figure*}[htb]
  \centering
  \includegraphics[width=10cm,height=8cm]{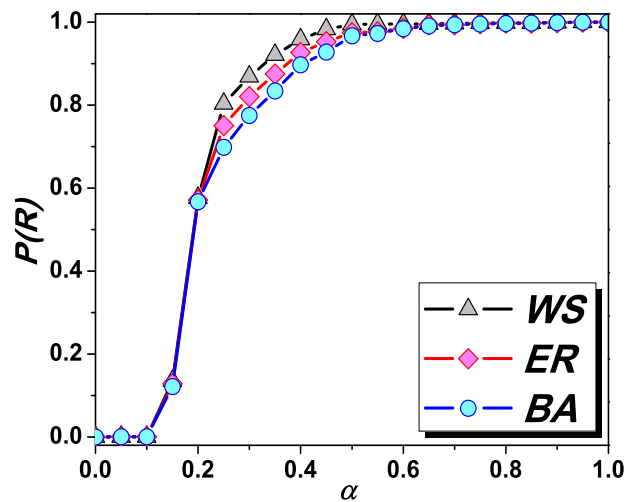}
  \caption{(Color online)\label{Fig:fig20} The ratio of exhausted nodes $P(R)$ as a function of $\alpha$ for three observed networks.}
\end{figure*}

Denote $P(R)$ as the fraction of exhausted nodes to the network size, obviously, larger $P(R)$ at the stable state indicates broader information spreading, and vice verse. To determine how the introduced parameter $\alpha$ affects the spreading results, we start our analysis from observing the relationship between $P(R)$ and $\alpha$. Figure \ref{Fig:fig20} shows that, for all the three networks, $P(R)$ is monotonically increasing with $\alpha$, suggesting a positive correlation  $P(R)$ and $\alpha$. As a consequence, in the following, we will use $P(R)$ instead of $\alpha$ to give comprehensive discussions.

\begin{figure*}[htb]
  \centering
  \includegraphics[width=18cm,height=6cm]{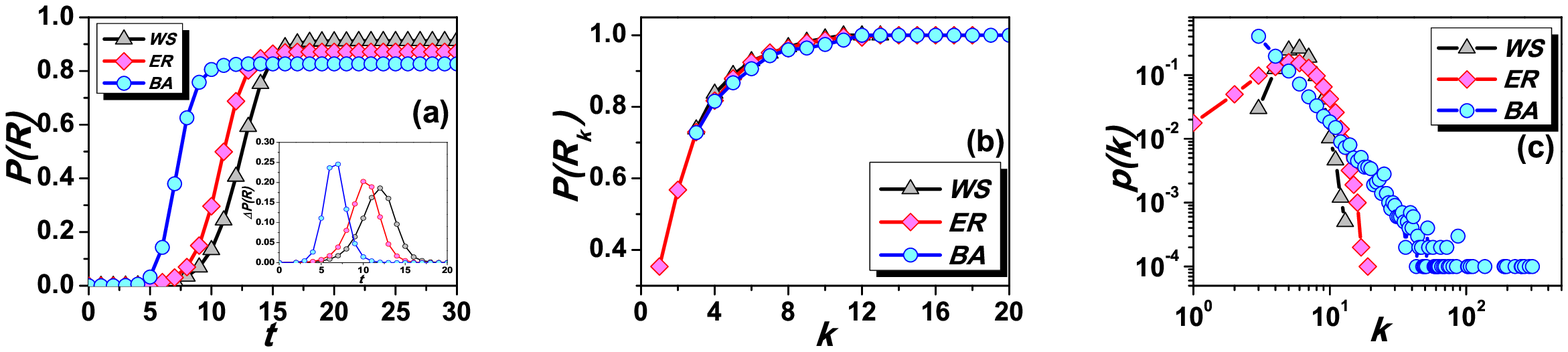}
  \caption{(Color online)\label{Fig:fig2} The dynamics of spreading on three networks. (a) Exhausted rate $P(R)$ as a function of simulation time. Inset is $\Delta P(R)$ as the function of simulation time; (b) Exhausted rate as a function of the node degree; (c) Degree distributions of three networks in log-log scale. The parameter $\alpha$ is set to 0.35 for (a) and (b). }
\end{figure*}

Figure \ref{Fig:fig2} reports the model results. In Figure \ref{Fig:fig2}a, it shows the dynamics of $P(R)$ on three different networks with the fixed parameter $\alpha=0.35$. It can be seen that WS network yields the most efficient spreading as the largest $P(R)$ at the stable state, while BA network exhibits the narrowest spreading, and ER network stays moderately. This result agrees with previous studies \cite{Lu-Chen-Zhou-2011-NJP}. In addition, the inset shows the dynamic difference of $P(R)$ between two continuous time steps, denoted as $\Delta P(R)=P(R_{t+1})-P(R_t)$, which can be considered as the spreading speed. It indicates that information spreads faster at its initial stage while slows down at later time steps due to the influence of hub nodes. From the degree distribution of Figure \ref{Fig:fig2}c, it can be seen that BA network is occupied by more large-degree nodes than other two. Therefore, information spreading on BA network is the fastest while slowest on WS network. According to the partial interaction in the present model, informed agents transmit information to only a certain fraction of their neighbors. Consequently, there would be also some uninformed individuals remaining without receiving any news because they are surrounded by exhausted neighbors (see Figure \ref{Fig:fig3}). Considering that the spreading process may influence differently to different type of nodes, we observe the relationship between exhausted rate $P(R_{k})$ and node degree $k$ of three networks at the stable state (see Figure \ref{Fig:fig2}b), where $P(R_{k})$ is the ratio of the total number of exhausted nodes with degree $k$ to the overall number of nodes with degree $k$. It clearly shows that $P(R_{k})$ and $k$ are apparently positively correlated, and nodes with smaller degree have less probability to receive information. Furthermore, Figure \ref{Fig:fig2}c displays the degree distributions of three networks. In general, BA network shows the power-law degree distribution where most nodes are of low degree but still a few hub nodes exist due to the \emph{rich-get-richer} mechanism. Poisson degree distribution emerges in ER network because of purely random attachment. Comparatively, most nodes in WS have moderate degree and only a few small and large degree nodes (but less than those in BA network), resulting from the rewiring process. In a word, the generally different fraction of hub nodes (Figure \ref{Fig:fig2}c) and different spreading effects on different type of nodes (Figure \ref{Fig:fig2}b) support the result of Figure \ref{Fig:fig2}a from the perspectives of structure and function, respectively.


\subsection{Information Blind Areas}
\label{Sec:Information Island}
\begin{figure}[htb]
  \centering
  \includegraphics[width=8cm,height=10cm]{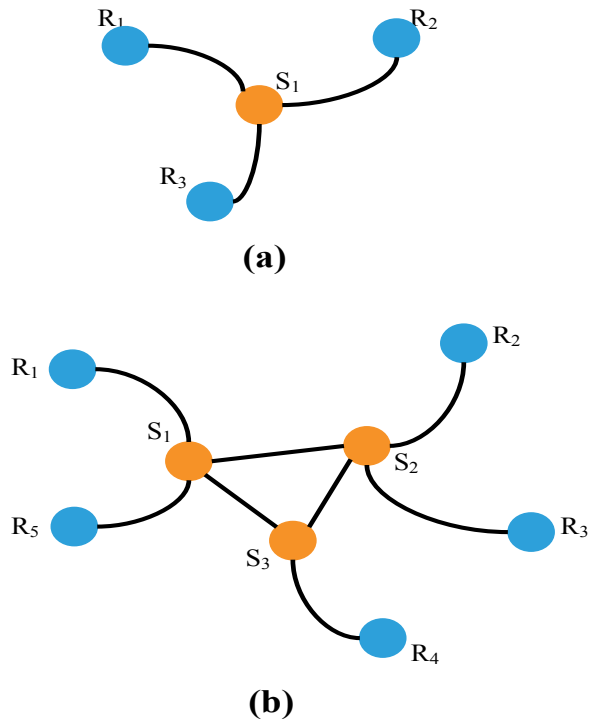}
  \caption{(Color online)\label{Fig:fig3} Illustration of two typical kinds of information blind areas: (a) one single-node case; and (b) example of multiple nodes.}
\end{figure}

As can be seen from Figure \ref{Fig:fig2}a, the exhausted rate is always less than 1, suggesting that some nodes will never be informed during the whole spreading process. Specifically, in this paper, we name the region consists of such unreachable nodes as the \emph{Information Blind Area}. Actually, Such blind areas emerge regularly due to the spreading process. Figure \ref{Fig:fig3} shows two typical kinds of information blind areas. One is composed by a single unformed node surrounded by exhausted neighbors and thus it could not receive news. As shown in Figure \ref{Fig:fig3}a, the uninformed node $S_{1}$ is enveloped by three exhausted nodes $R_{1}$, $R_{2}$ and $R_{3}$, who would neither directly transmit information to $S_{1}$ nor allow the informed agents to contact  $S_{1}$. The other one is composed of multiple uninformed nodes connected with each other and surrounded together by exhausted nodes. Thus, the information will never be transmitted to such a \emph{cluster} as no path is available. As shown in Figure \ref{Fig:fig3}b, three uninformed nodes, $S_{1}$, $S_{2}$ and $S_{3}$, compose a connected cluster which is surrounded by five exhausted nodes, $R_{1}$, $R_{2}$, $R_{3}$, $R_{4}$ and $R_{5}$. Since no
single uniformed node in this cluster can access the outside information world, they have to passively keep \emph{blind} for the spreading information.

\begin{figure*}[htb]
  \centering
  \includegraphics[width=16cm,height=16cm,angle=90]{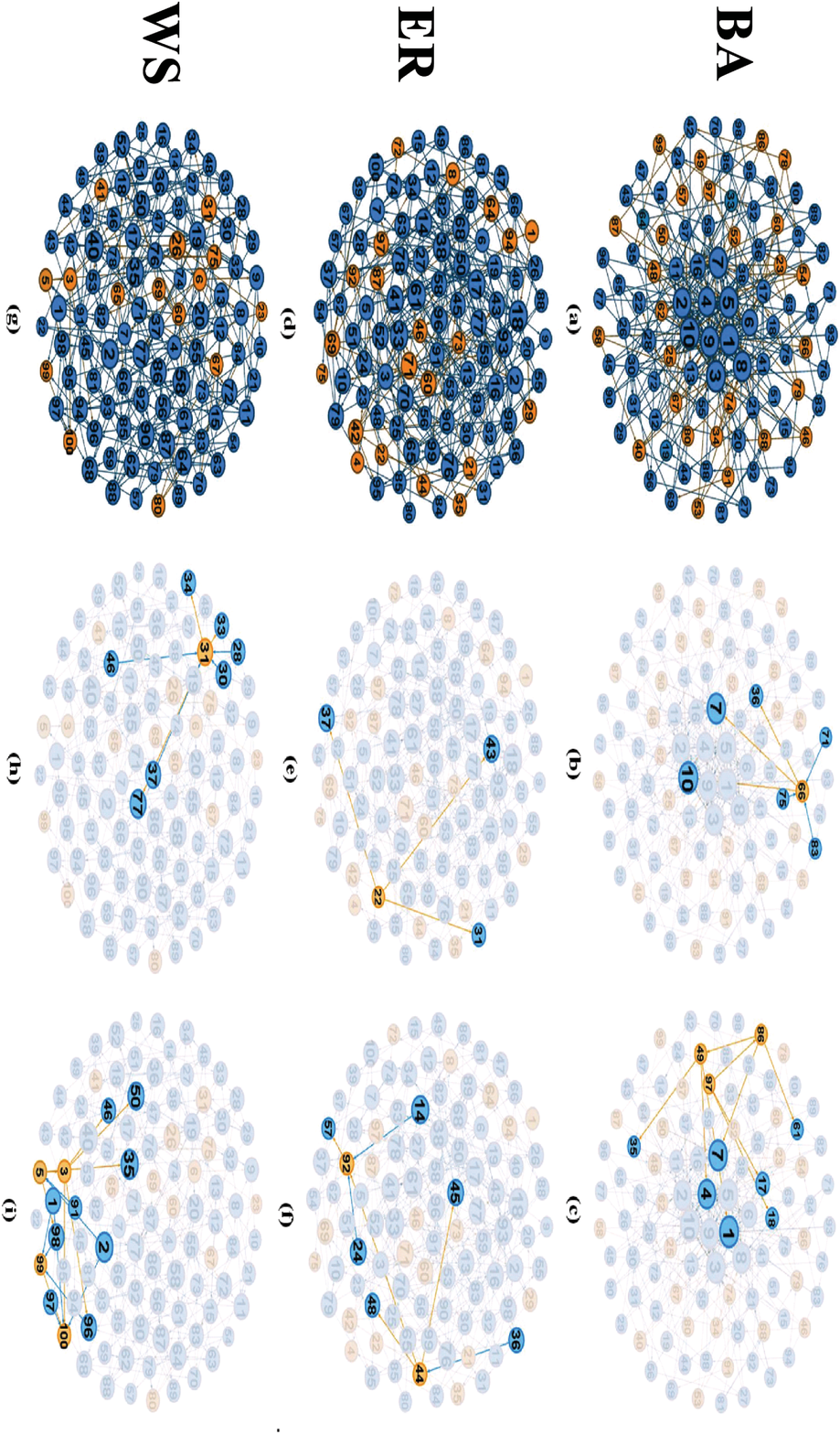}
  \caption{(Color online)\label{Fig:fig4} Visualization of information blind areas resulted from designed experiments. Yellow and blue circles represent the uninformed and exhausted nodes, respectively. The size of each node is proportional to its degree. In addition, (a-c) are the results on BA network; (d-f) are the results on ER network; (g-i) are the results on WS network. }
\end{figure*}

In order to describe those two kinds of information blind areas more clearly, we conduct experiments with $N$=100 and $\langle k \rangle$=6. Figure \ref{Fig:fig4} shows the visualization results on the three representative networks. It can be seen that most uninformed individuals at stable state (Figure \ref{Fig:fig4}a, \ref{Fig:fig4}d, \ref{Fig:fig4}g) are generally small-degree ones, which is consistent with previous analysis in Section \ref{Sec:Exhausted Rate}. In addition, we also show the two information blind areas. The single-node case is displayed in Figure \ref{Fig:fig4}b, Figure \ref{Fig:fig4}e, Figure \ref{Fig:fig4}h. In Figure \ref{Fig:fig4}b, the uninformed individual \#66 is surrounded by exhausted neighbors \#7, \#10, \#36, \#71, \#75 and \#83. As a consequence, node \#66 is prevented from hearing news by its neighbors. Figure \ref{Fig:fig4}e and Figure \ref{Fig:fig4}h are similar to that of Figure \ref{Fig:fig4}b. On the other hand, Figure \ref{Fig:fig4}c, \ref{Fig:fig4}f, and \ref{Fig:fig4}i report the information blind areas of multi-nodes cases on BA, ER and WS networks, respectively. As shown in Figure \ref{Fig:fig4}c, three uninformed individuals \#49, \#86 and \#97 compose a connected cluster, which is surrounded by exhausted individuals \#1, \#4, \#7, \#17, \#18, \#35, and \#61. Therefore, all the nodes in this relatively large area will not be informed and keep blind to spread information. Similar phenomena can also be discovered in Figure \ref{Fig:fig4}f and Figure \ref{Fig:fig4}i on ER and WS networks, respectively.

\subsection{Scale-free Effect}
\label{Sec:SF}
\begin{figure*}[htb]
  \centering
  \includegraphics[width=18cm,height=14.5cm]{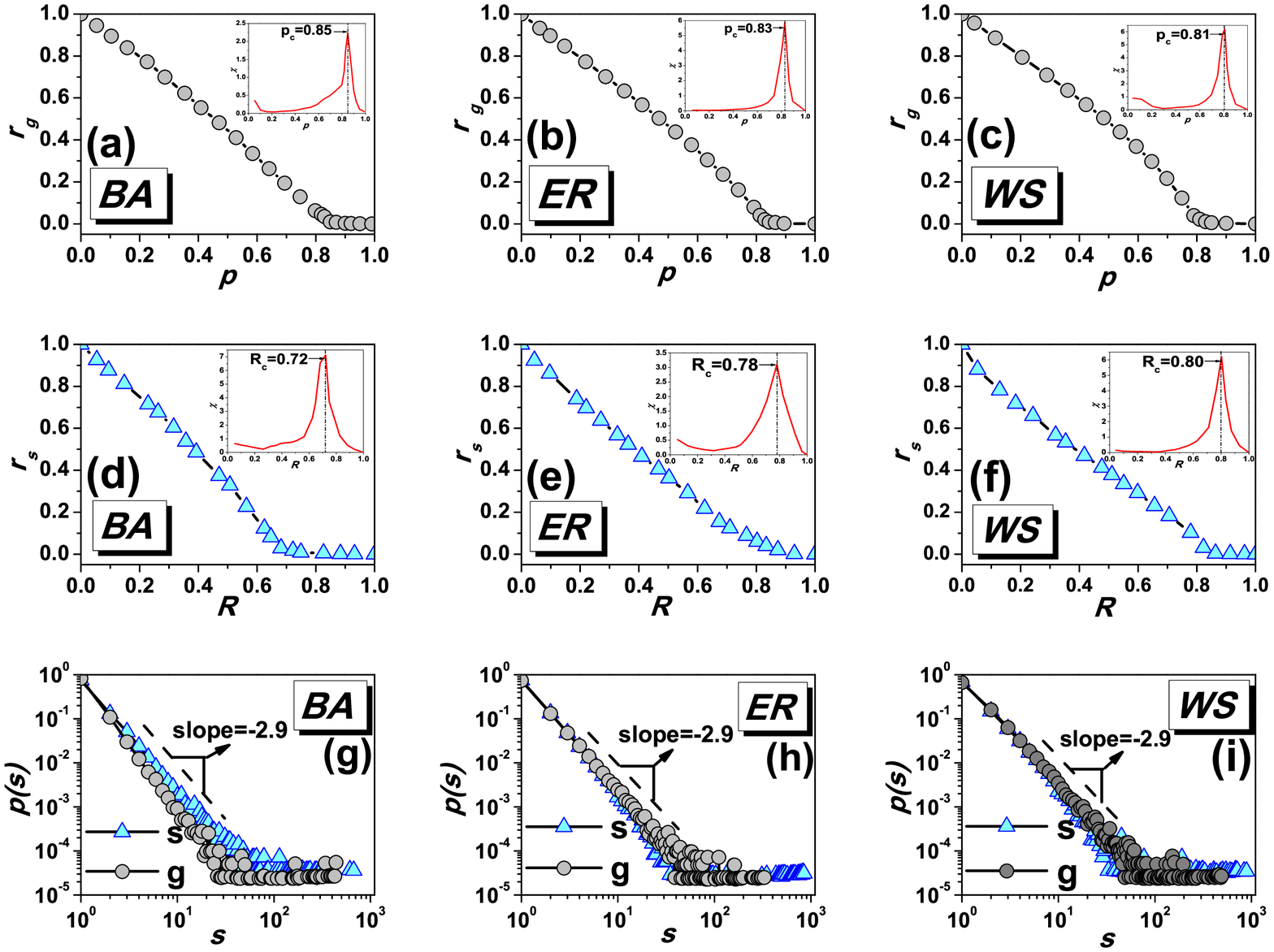}
  \caption{(Color online)\label{Fig:fig5} Comparisons between the present model and the site percolation system. (a-c) The relationship between the size of giant cluster and empty probability $p$. Inset shows $\chi$ versus $p$, and the peak is the critical point $p_c$; (d-f) The relationship between the size of the largest information blind areas and the exhausted rate $R$ .Inset shows $chi$ versus $R$, and the peak is the critical point $R_c$; (g-i) The size distributions of connected clusters of the two models.}
\end{figure*}

In this section, we shall explain how such blind areas emerge from the perspective of site percolation.	
Generally, in a typical site percolation system, each node in the given network would be at one of two given states: \emph{empty} with probability $p$ or \emph{occupied} with probability $1-p$, and all the edges connecting with empty nodes are cut off.
Previous studies have proved that the site percolation system would display a phase transition
phenomenon from \emph{connected} phase to \emph{disconnected} phase when $p$ grows to a critical point $p_c$ \cite{Dorogovtsev2008RMP}.
At this point, the behavior of the ratio of giant cluster size (the largest connected cluster) $r_g$ ($r_g = N_g/N$, where $N_g$ is the size of the giant cluster) indicates the network's sudden disintegration, and the scaling distribution of connected clusters would emerge \cite{newman1999scaling,cohen2000resilience,albert2000error,cohen2002percolation,schwartz2002percolation}.
Figure \ref{Fig:fig5}a-\ref{Fig:fig5}c respectively show how $r_g$ changes according to different $p$, as well as the critical point (see insets) in three different networks (BA, ER and SW networks) with the same network size $N$=10000.


Note that, the role of the empty nodes in site percolation is quite similar with that of the R-state individuals in the present model,
where the information blind areas are divided by those individuals. Analogous with observing the giant cluster in site percolation,
we consequently calculate the ratio of the largest blind area size $r_s$ ($r_s = N_s/N$, where $N_s$ is the size of the largest information blind area) for various exhausted rate $R \in [0,1]$.
The results are shown in Figure \ref{Fig:fig5}d-\ref{Fig:fig5}f, where the curve exhibits a very similar trend with that of $r_g$.
In addition, we numerically calculate the thresholds of both $p_c$ and $R_c$, according to the peak of the relative variance of the size distribution.
We denote it by $\chi = \frac{\langle N_c^2 \rangle - \langle N_c \rangle^2}{\langle N_c \rangle}$, where $N_c$ is the size of each connected cluster except the giant one \cite{ferreira2012epidemic}, for it would result in large fluctuation when $p$ or $R$ approaches the critical point, and the insets of Figure \ref{Fig:fig5}a-\ref{Fig:fig5}f show the corresponding results.
Furthermore, in different networks (BA, ER and SW), we find good agreements between the size distributions of connected clusters and information blind areas
of the two respective modes. Figure \ref{Fig:fig5}g-\ref{Fig:fig5}i show that the distributions of the connected clusters' size of the two models are coincident with almost the same exponent -2.9 at their respective critical points, indicating that they have very similar critical phenomena.

Despite those similar properties, we should not regard the present model as an equivalent process as the site percolation.
One significant difference between them is that, the node states are changed according to the diffusion process based on the given network structure in the spreading model, while the site percolation process only randomly label the node states regardless of the network structure. Therefore, the critical points exhibit very  differently for the two models. In the insets of Figure \ref{Fig:fig5}a-\ref{Fig:fig5}c, the critical values follow $p^{BA}_c > p^{ER}_c > p^{WS}_c$ for the site percolation, while a completely opposite sequence for the proposed model as shown in the insets of Figure \ref{Fig:fig5}d-\ref{Fig:fig5}f (that is $R^{BA}_c < R^{ER}_c < R^{WS}_c$). For the random labeling process, the more heterogeneous the network is, the larger the critical point will be, as there are more small-degree nodes when each one is treated equally. On the contrary, if there exist some large-degree nodes dominating the network center, messages can be easily spread out via those core nodes, hence the critical point will be smaller for such large heterogeneous network in the diffusion process. Similar statements can also be applied in illustrating that, for the same network, the critical value of the interaction model is always smaller than that of site percolation, e.g. $p^{BA}_c > R^{BA}_c$, $p^{ER}_c>R^{ER}_c$ and $p^{WS}_c > R^{WS}_c$. In a word, although the site percolation principle can not be fully projected to spreading dynamics, it still provides a promising and versatile tool to explain the critical phenomenon of the emergence of blind areas in information spreading.


\section{Conclusions \& Discussion}
\label{Sec:Conclusion}
In this paper, we have applied the partial interaction effect in the classical SIR model where three type of node states are considered: (1) Uninformed (S); (2) Informed (I); (3) Exhausted (R). Subsequently, we adopt it in an information spreading scenario, where the
interaction strength is proportional to each spreader's degree. Numerical experiments show that there is a clearly positive relationship between
the interaction strength and final coverage of spreading. In addition, we find that the spreading effect is highly influenced by the network structure. One interesting property will result from the spreading process, where the network is divided into several information-unreachable areas, namely \emph{Information Blind Areas}. Analysis reveals that a phase transition of such blind areas' size distribution will emerge when the fraction of R-state individuals grow to a certain critical point. In spite of quite similar results from site percolation analysis, detailed experiments show that the diffusion process is significantly different from random selection. Further numerical analyses on three representative networks (BA, ER and WS) demonstrate that, for the spreading process, the more heterogeneous the network is, the smaller the critical point will be. By contrast, the random labeling process of site percolation is completely opposite. Moreover, the critical value of network-based diffusion is smaller than that of purely random selection. Those findings can be regarded as additional explanation why small-world network yields the most efficient information/epdemic spreading in previous studies \cite{Keeling-Eames-2005-JRSI, Lu-Chen-Zhou-2011-NJP}.

Recently, the research of both contagion-based spreading models and applications has attracted more and more
attention \cite{Karsai-Kivela-Pan-Kaski-Kertesz-Barabasi-Saramaki-2011-PRE,Iribarren-Moro-2011-PRE}. Numerical results in this paper demonstrate that, due to the various network structure, there are always unreachable individuals. Human communication pattern analysis \cite{Miritello-Moro-Lara-2011-PRE, JiangZQ2013} would be a promising method to help in understanding how to lighten the \emph{Dark Corners}. In addition, activation of \emph{Long-tail} individuals and products would also enhance the efficiency of \emph{Information Filtering} \cite{Lv2012} in the era of big data to solve the \emph{cold-start} dilemma \cite{qiu2011EPL, qiu2013PO}. 

\section*{Acknowledgments}

This work was partially supported by the National Natural Science Foundation of China (Grant Nos. 11105024, 11105040, 1147015, 11301490 and 11305043),
the Zhejiang Provincial Natural Science Foundation of China (Grant Nos. LY12A05003 and LQ13F030015), the EU FP7 Grant 611272 (project GROWTHCOM),
the start-up foundation and Pandeng project of Hangzhou Normal University.

\section*{References}
\bibliographystyle{iopart-num}

\bibliography{Bibliography}

\end{document}